\documentclass[aps,pra,twocolumn,superscriptaddress,groupedaddress]{revtex4}

\usepackage{amssymb} \usepackage{color,graphicx} \usepackage{amsmath}
\usepackage{amsbsy} \usepackage{amsthm} \usepackage{bbm}
\usepackage{bm,bbm} \usepackage{float} \usepackage{braket}
\usepackage{placeins}
\usepackage[colorlinks=true,citecolor=blue,linkcolor=red,urlcolor=red]{hyperref}
\usepackage[sort&compress]{natbib}
\usepackage{comment}
\usepackage{stackengine}

\newcommand{\bq}{\begin{equation}} \newcommand{\eq}{\end{equation}}
\newcommand{\bqali}{\bq\begin{aligned}}
\newcommand{\eqali}{\end{aligned}\eq}
\newcommand{\bqn}{\begin{equation*}}
\newcommand{\eqn}{\end{equation*}}

\newcommand\D{\operatorname{d}\!}
\renewcommand\k{{\bf k}}

\newcommand\z{{\bf z}}

\newcommand\x{{\bf x}}
\newcommand\y{{\bf y}}

\newcommand\kb{k_\text{\tiny B}}

\newcommand\erf{\operatorname{erf}}
\newcommand\erfc{\operatorname{erfc}}
\newcommand\com[2]{[#1,#2]}
\newcommand\acom[2]{\{#1,#2\}}

\newcommand\omegam{\omega_\text{\tiny m}}

\newcommand\gammam{\gamma_\text{\tiny m}}

\newcommand\DNSff{\mathcal S_{\text{\tiny CSL}}}
\newcommand\DNS{\mathcal S(\omega)}
\newcommand\DNSc{\mathcal S_\text{\tiny cCSL}(\omega)}

\newcommand\q{{\bf q}}
\newcommand\rC{r_\text{\tiny C}}

\graphicspath{{../Figures/}}

\begin{document}

\author{Matteo Carlesso}
\email{matteo.carlesso@ts.infn.it}
\affiliation{Department of Physics, University of Trieste, Strada Costiera 11, 34151 Trieste, Italy}
\affiliation{Istituto Nazionale di Fisica Nucleare, Trieste Section, Via Valerio 2, 34127 Trieste, Italy}

\author{Luca Ferialdi}
\email{l.ferialdi@soton.ac.uk}
\affiliation{Department of Physics and Astronomy, University of Southampton, SO17 1BJ, UK}

\author{Angelo Bassi}
\affiliation{Department of Physics, University of Trieste, Strada Costiera 11, 34151 Trieste, Italy}
\affiliation{Istituto Nazionale di Fisica Nucleare, Trieste Section, Via Valerio 2, 34127 Trieste, Italy}

\title{{Colored collapse models from the non-interferometric perspective}}

\date{\today}
\begin{abstract}

Models of spontaneous wave function collapse  describe the quantum-to-classical transition by assuming a progressive breakdown of the superposition principle when the mass of the system increases, providing a well-defined phenomenology in terms of a non-linearly and stochastically modified Schr\"odinger equation, which can be tested experimentally. The most popular of such models is the Continuous Spontaneous Localization (CSL) model: in its original version, the collapse is driven by a white noise, and more recently, generalizations in terms of colored noises, which are more realistic, have been formulated. We will analyze how current non-interferometric tests bound the model, depending on the spectrum of the noise. We will find that  low frequency purely mechanical experiments provide the most stable and strongest bounds.

\end{abstract}
 \maketitle

\section{Introduction}

Since the birth of Quantum Mechanics with its striking differences compared to our classical intuition, the quantum-to-classical transition has puzzled the scientific community. The scientific debate, first confined to conceptual arguments,  now has an experimental counterpart, thanks to technological developments, which allow to put at test the questions about  the boundaries between the classical and quantum realms \cite{Joos:1985aa,Zurek:1991aa,Joos:2003aa,Schlosshauer:2007aa}.

The quantum-to-classical transition is consistently described in terms of collapse models \cite{Bassi:2003aa,Bassi:2010aa}. These are phenomenological models, which modify the standard Schr\"odinger dynamics by introducing new suitable non-linear and stochastic terms, which properly describe the collapse of the wave-function.

The Continuous Spontaneous Localization (CSL) model \cite{Ghirardi:1990aa} is the most studied among collapse models, and nowadays it represents a commonly accepted alternative to standard quantum mechanics. The stochastic terms characterizing this model depend on two parameters: the collapse rate $\lambda$ and the noise correlation length $\rC$. 
Based on alternative theoretical considerations, different numerical values have been proposed: $\lambda=10^{-16}\,$s$^{-1}$ and $\rC=10^{-7}\,$m by Ghirardi, Rimini and Weber \cite{Ghirardi:1986aa}; $\lambda=10^{-8\pm2}\,$s$^{-1}$ for $\rC=10^{-7}\,$m, and $\lambda=10^{-6\pm2}\,$s$^{-1}$ for $\rC=10^{-6}\,$m by Adler \cite{Adler:2007ab}.
However, since the CSL model is phenomenological, at present the values of the parameters can only be bounded by experiments. 

Although {only recently the scientific community has started developing dedicated experiments \cite{Vinante:2017aa,Bassi:2018aa}}, one can infer bounds on the CSL parameters by investigating the predictions of the model for suitable experimental scenarios. One can divide them in two classes: interferometric experiments and non-interferometric ones. The first class includes those experiments where ideally a spatial superposition  is created in an interferometer, and the corresponding interference pattern is measured. This is the case of cold-atom \cite{Kovachy:2015aa} and molecular~\cite{Eibenberger:2013aa,Hornberger:2004aa,Toros:2017aa,Toros:2018aa} interferometry, and entanglement experiments with diamonds \cite{Lee:2011aa,Belli:2016aa}. Conversely, the experiments falling in the second class are those where no superposition is generated,  the collapse being detected indirectly through the random motion which is always associated to it. These experiments involve cold atoms~\cite{Kovachy:2015ab}, optomechanical systems~\cite{Usenko:2011aa,Vinante:2017aa,Vinante:2006aa,Abbott:2016ab,Abbott:2016aa,Armano:2016aa,Armano:2018aa}, X-ray measurements~\cite{Aalseth:1999aa}, phonon excitations in crystals~\cite{Adler:2018aa,Bahrami:2018aa}. Note that in non-interferometric experiments one can also consider systems which are (truly) macroscopic. In such a case, due to the amplification mechanism, the collapse can be more significant and easier to detect. This has been shown for example in \cite{Vinante:2016aa,Vinante:2017aa} for  a micrometer cantilever, and in \cite{Carlesso:2016ac,Helou:2017aa} for human-scale gravitational wave detectors: these experiments establish the strongest bounds on $\lambda$ for $\rC>10^{-6}\,$m, while X-ray measurements~\cite{Aalseth:1999aa}, which also employ human-scale objects, set the strongest bounds for  $\rC<10^{-6}\,$m.

Although CSL well describes the collapse process, it has two weak points.
The first one is that the interaction with the collapse noise heats up the system of interest, making its average energy increase linearly. Although such a heating  has rather long time-scales (making it negligible in most situations), this is something one eventually would like  to remove.  This has been resolved by the dissipative extension of the the model~\cite{Bassi:2003aa,Smirne:2014aa,Smirne:2015aa,Toros:2018aa,Toros:2017aa,Nobakht:2018aa}.

The second weak point concerns the spectrum of the CSL noise, which is flat, being the noise white. If one thinks that the noise providing the collapse has a physical origin, it cannot be white but colored, with a cut off. This extension of the CSL model has already been formulated~\cite{Adler:2007aa,{Adler:2008aa},Bassi:2009aa,Ferialdi:2012aa,Adler:2013aa,Bilardello:2016aa,Toros:2017aa}, and we will refer to it as  ``colored CSL'' (cCSL): it the subject of the present article. In particular, we will investigate the bounds that non-interferometric experiments place on the collapse parameters $\lambda$ and $\rC$, when a colored noise with exponentially decaying correlation function is considered.\\

The paper is organized as follows: in Sec.~\ref{sec.2} we introduce the cCSL model and we compute its contribution to the density noise spectrum of optomechanical systems. In Sec.~\ref{sec.3} we quickly present the cCLS predictions for other relevant non interferometric experiments: X-ray measurements \cite{Piscicchia:2017aa} and bulk heating \cite{Adler:2018aa,Bahrami:2018aa}. In Sec.~\ref{sec.4} we use these theoretical formulas to derive the bounds on the cCSL parameters from available experimental data, and in Sec.~\ref{sec.5} we discuss the results and draw our conclusions.

\section{CSL model and optomechanical systems}
\label{sec.2}
Before digging into the details of the cCSL model, we start by reviewing the basic features of CSL model, which will be useful for the following analysis. As already briefly described before, the CSL model modifies the standard Schr\"odinger dynamics for the wave-function by adding to it non-linear and stochastic terms \cite{Bassi:2003aa}.
Non-linearity is required in order to suppress quantum superpositions, while stochasticitiy has to be implemented in order to  avoid superluminal signaling and to recover the Born rule in measurement situations \cite{Bassi:2003aa}.
These modifications are devised in such a way that the motion of microscopic objects is not significantly affected by them (hence recovering standard quantum mechanics), while an in-built amplification mechanics guarantees that macroscopic bodies behave classically.\\

A non-linear stochastic collapse equation is rather difficult to solve. However, when one comes to the expectation values of physical quantities, the  CSL collapse can be mimicked by the  random potential~\cite{Carlesso:2016ac}
\bq\label{CSLpotential}
\hat V_\text{\tiny cCSL}(t)=-\frac{\hbar\sqrt{\lambda}}{m_0}\int\D\z\, \hat M(\z)w(\z,t),
\eq
with $w(\z,t)$ a classical Gaussian noise characterized by
\bq\label{whitecsl}
\mathbb E[{w(\z,t)}]=0, \quad\mathbb E[{w(\z,t)w(\x,s)}]=\delta^{(3)}(\z-\x)f(t-s),
\eq
where $\mathbb E[\ \cdot\ ]$ denotes the stochastic average. { In the general case, $w(\z,t)$ has a correlation function $f(t)$ with a non-trivial (colored) spectrum. In particular, the standard CSL model is recovered with $f(t)=\delta(t)$,  implying a flat spectrum (white noise).}
In Eq.~\eqref{CSLpotential} we introduced 
$\hat M(\z)$, which is a locally averaged mass density operator:
\bq\label{def1M}
\hat M(\z)=\frac{m_0}{\pi^{3/4}\rC^{3/2}}\sum_ne^{-\tfrac{(\z-\hat\q_n)^2}{2\rC^2}},
\eq
where $\hat \q_n$ is the position operator of the $n$-th nucleon of the system and $m_0$ is a reference mass chosen equal to the one of a nucleon. 
When the spread in position of the center of mass is much smaller than $\rC$, under the rigid body assumption, we can approximate the above expression with \cite{Nimmrichter:2014aa}
\bq\label{exp.M}
\hat M(\z)\simeq M_0(\z)+\int\D\x\,\frac{\mu(\x)}{{\pi^{3/4}\rC^{7/2}}}e^{-\tfrac{(\z-\x)^2}{2\rC^2}}(\z-\x)\cdot\hat\q,
\eq
where $M_0(\z)$ is a complex function, $\mu(\x)$ is the mass density of the system and $\hat \q$ is the center of mass operator.

We now have all the key ingredients to evaluate the effects of the CSL model on optomechanical systems~\cite{Nimmrichter:2014aa,Bahrami:2014aa,Diosi:2015ab}. Their dynamics  is conveniently described in terms of a Langevin equation, which we here write down in its one dimensional version (along the $x$ direction), in the limit of vanishing optical coupling \cite{Carlesso:2018ab}:
\bq\label{langevin.eq}
\frac{\D{\hat x}}{\D t}=\frac{\hat p}{m}, \quad\frac{\D{\hat p}}{\D t}=-m\omegam^2\hat x-\gammam \hat p+\hat \xi(t)+F_\text{\tiny cCSL}(t).
\eq
Here, $m$ is the mass of the system, $\omegam$ is the frequency of the harmonic trap and $\gammam$ is the damping constant. We introduced two stochastic terms of different origin: $\hat \xi(t)$ and $F_\text{\tiny cCSL}(t)$. The former is quantum, and describes the thermal action of the surrounding environment, which is supposed to be in equilibrium at temperature $T$. Its average and correlation are $\braket{\hat \xi(t)}=0$ and
\bq
\braket{\hat \xi(t)\hat \xi(s)}=\hbar m\gammam\int\frac{\D\omega}{2\pi}\,e^{-i\omega(t-s)}\omega\left[1+\coth\left(\tfrac{\hbar\omega}{2\kb T}\right)\right].
\eq
The second stochastic contribution is classical and describes the CSL action on the system, according to $F_\text{\tiny cCSL}(t)=\tfrac{i}{\hbar}\com{\hat V_\text{\tiny cCSL}(t)}{\hat p}$. Given the form of $\hat V_\text{\tiny cCSL}(t)$ and $\hat M(\z)$ respectively in Eq.~\eqref{CSLpotential} and Eq.~\eqref{exp.M}, we find that \cite{Carlesso:2016ac}
\bq\label{forceCSL}
F_\text{\tiny cCSL}(t)=\frac{\hbar \sqrt{\lambda}}{\pi^{3/4}m_0}\int\D\z\D\x\,\frac{\mu(\x)}{\rC^{7/2}}e^{-\tfrac{(\z-\x)^2}{2\rC^2}}(\z-\x)_xw(\z,t).
\eq
We note that, with reference to Eq.~\eqref{def1M}, without approximations the stochastic force would be an operator. However, since in the Taylor expansion of Eq.~\eqref{exp.M} we considered only the terms up to the first order in the center of mass position operator, $F_\text{\tiny cCSL}(t)$ becomes  a function (we restrict to this case throughout the paper).

Equation~\eqref{langevin.eq} allows us to derive the density noise spectrum, which  quantifies the overall noise on the system.

\subsection{Density Noise Spectrum}

The Density Noise Spectrum (DNS) $\DNS$ characterizes the motion of an optomechanical system in its steady state \cite{Gardiner:2004aa}. It is defined as the Fourier transform of the fluctuation $\delta\hat x(t)=\hat x(t)-\hat x_\text{\tiny SS}$ of the position operator   around its steady state  $\hat x_\text{\tiny SS}$:
$\DNS=\frac12\int_{-\infty}^{+\infty}\D\tau\,e^{-i\omega\tau}\mathbb E\left[\braket{\acom{\delta\hat x(t)}{\delta\hat x(t+\tau)}}	\right],
$ 
where $\braket{\ \cdot\ }$ denotes the quantum expectation value. The DNS can also be written in terms of $\delta\tilde x(\omega)$, which is the Fourier transform of $\delta\hat x(t)$:
\bq\label{defDNS}
\DNS=\frac{1}{4\pi}\int_{-\infty}^{+\infty}\D\Omega\,\mathbb E\left[\braket{\acom{\delta\tilde x(\omega)}{\delta\tilde x(\Omega)}}	\right].
\eq
Following the standard prescription \cite{Gardiner:2004aa,Paternostro:2006aa}, one can derive from Eq.~\eqref{langevin.eq} the equations of motion of $\delta\hat x(t)$ and of the similarly defined momentum fluctuations $\delta\hat p(t)=\hat p(t)-\hat p_\text{\tiny SS}$. By solving these equations in the Fourier space, we get:
\bq
\delta \tilde x(\omega)=\frac1m\frac{\tilde \xi(\omega)+\tilde F_\text{\tiny cCSL}(\omega)}{(\omegam^2-\omega^2)-i\gammam\omega},
\eq
where as general notation rule, we denote with tilde the Fourier transform of a quantity.
The correlation functions of the noises, in Fourier space, read:
\bqali
&\braket{\tilde \xi(\omega)\tilde \xi(\Omega)}=2\pi\hbar m\gammam\omega\left[1+\coth\left(\tfrac{\hbar\omega}{2\kb T}\right)\right]\delta(\omega+\Omega),\\
&\mathbb E\left[\acom{\tilde F_\text{\tiny cCSL}(\omega)}{\tilde F_\text{\tiny cCSL}(\Omega)}\right]=2\pi\hbar^2\eta\tilde f(\omega)\delta(\omega+\Omega),
\eqali
where $\tilde f(\omega)$ is the Fourier transform of $f(t)$ and
\bq\label{eta}
\eta=\frac{\lambda\rC^3}{\pi^{3/2}m_0^2}\int\D\k|\tilde \mu(\k)|^2 k_x^2e^{-\k^2\rC^2},
\eq
where $\tilde \mu(\k)$ is the Fourier transform of the mass density.
{ By exploiting Eqs.~\eqref{defDNS}-\eqref{eta} to evaluate the DNS, in the high temperature approximation we obtain the following identity:}
\bq\label{dns.eq}
\DNS=\frac{	2 m \gammam\kb T +\mathcal{S}_\text{\tiny cCSL}(\omega)}{m^2[(\omegam^2-\omega^2)^2+\gammam^2\omega^2]},
\eq
where $\mathcal{S}_\text{\tiny cCSL}(\omega)=\tfrac1{4\pi}\int\D\Omega\,\mathbb E[\acom{\tilde F_\text{\tiny cCSL}(\omega)}{\tilde F_\text{\tiny cCSL}(\Omega)}]$ is the CSL contribution to the DNS.

In the standard CSL model,  whose noise is white, one gets:
\bq\label{dnscsl}
\DNSff=\hbar^2\frac{\lambda\rC^3}{\pi^{3/2}m_0^2}\int\D\k|\tilde \mu(\k)|^2 k_x^2e^{-\k^2\rC^2},
\eq
which is independent of $\omega$, being the noise spectrum flat.
Conversely, in its colored extension, one has:
\bq
\DNSc=\DNSff \times\tilde f(\omega).
\eq
The frequency dependent contribution due to the noise correlation function potentially can lead to relevant modifications of the bounds on the collapse parameters $\lambda$ and $\rC$. 

\section{Other non-interferometric tests}
\label{sec.3}

We briefly report the theoretical analysis of other two significant non-interferometric tests of collapse models:  X-ray measurements \cite{Piscicchia:2017aa} and low temperature measurements {of phonon vibrations} \cite{Adler:2018aa,Bahrami:2018aa}, which we will use to infer the bounds on the cCSL parameters. Also cold atom experiments \cite{Kovachy:2015ab} are significant for testing CSL; as they have been fully analyzed in~\cite{{Bilardello:2016aa}}, we refer to that paper for their derivation, and we will simply report the result in the following section.\\

\noindent \textit{A. X-ray emission.} In  testing CSL with X-ray measurements~\cite{Fu:1997aa,Adler:2007ad,Adler:2013aa,Bassi:2014aa,Donadi:2014aa,Piscicchia:2017aa}, the basic idea is that electrons and protons in the sample material are accelerated by the collapse noise and  emit radiation, which can be detected. Under suitable approximations the associated photon emission rate reads \cite{Nobakht:2018aa}
\bq
\frac{\D \Gamma(\omega)}{\D\omega}=\frac{e^2\hbar\eta}{2\pi^2\epsilon_0 c^3m_\text{\tiny e}^2}\frac{1}{\omega}\tilde f(\omega),
\eq
where $e$ is the unitary charge, $\epsilon_0$ the dielectric constant of vacuum and $c$ the speed of light, $m_\text{\tiny e}$ is the electron mass and $\eta=\lambda m_\text{\tiny e}^2/(2m_0^2\rC^2)$. The standard CSL expression is obtained by setting $\tilde f(\omega)=1$.\\

\noindent \textit{B. Phonon excitation.} Recently a novel way to test  CSL was proposed, setting strong bounds on $\lambda$ for $\rC< 10^{-6}$\,m \cite{Adler:2018aa,Bahrami:2018aa}. The idea is that the CSL noise modifies the phonon's spectrum in a material, while heating it. This modification is quantified by the energy gain rate per mass unit:
\bq\label{Erate}
\frac{\D E}{\D t\D M}=\frac{3}{4}\frac{\hbar^2}{\rC^2m_0^2}\lambda_\text{\tiny eff},
\eq
where 
\bq\label{Lphonon}
\lambda_\text{\tiny eff}=\frac{2\lambda\rC^5}{3\pi^{3/2}}\int\D\q\,e^{-\q^2\rC^2}\q^2\tilde f(\omega_\text{\tiny L}(\q)).
\eq
with $\omega_\text{\tiny L}(\q)$ denoting the longitudinal phonon frequency, which explicitly depends on the momentum $\q$.
In the white noise case $\lambda_\text{\tiny eff}=\lambda$, recovering the standard CSL result for the energy heating~\cite{Adler:2018aa}.

\section{Experimental bounds}
\label{sec.4}

Having set all theoretical formulas, we pass now to deriving the bounds on the cCSL parameters from available experimental data. As an explicit example, we consider an exponentially decaying noise correlation function with correlation time $\Omega_\text{\tiny c}^{-1}$:
\bq\label{f.time.corr}
f(t-s)=\frac{\Omega_\text{\tiny c}}{2}e^{-\Omega_\text{\tiny c}|t-s|}.
\eq
Such a choice, besides being physically reasonable, allows to easily recover the white noise limit for $\Omega_\text{\tiny c}\rightarrow\infty$.
This form of the correlation function was already considered in \cite{Bassi:2009aa,Bilardello:2016aa} in the context of colored modifications of collapse models.

For the optomechanical setting discussed in Sec.~\ref{sec.2}, our choice of noise corresponds to $\DNSc$ in the Drude-Lorentz form
\bq\label{dnsc}
\DNSc=\DNSff\frac{\Omega_\text{\tiny c}^2}{\Omega_\text{\tiny c}^2+\omega^2},
\eq
where we see that $\Omega_\text{\tiny c}$ plays the role of frequency cutoff on the noise spectrum. \\

In a similar way, we can straightforwardly derive the photon emission rate for this specific time correlation function of the cCSL noise:
\bq\label{gammaXray}
\frac{\D \Gamma(\omega)}{\D\omega}=\frac{e^2\hbar\eta}{2\pi^2\epsilon_0 c^3m_\text{\tiny e}^2}\frac{1}{\omega}\frac{\Omega_\text{\tiny c}^2}{\Omega_\text{\tiny c}^2+\omega^2}.
\eq
Again, in the limit of $\Omega_\text{\tiny c}\to+\infty$ we recover the expression for the standard CSL.\\

As for CSL induced phonon excitation, having a colored noise leads to more involved modifications of the energy gain rate, as the frequency $\omega_\text{\tiny L}(\q)$ appearing in Eq.~\eqref{Lphonon} in general depends in a non trivial way on the momentum. For a monoatomic crystal, the dispersion relation is~\cite{Kittel:2005aa}
\bq
\omega_\text{\tiny L}^2(\q)=\frac{4C}{m_\text{\tiny A}}\sin^2\left(\tfrac12|\q|a\right),
\eq
where $m_\text{\tiny A}$ is the atomic mass, $C$ is the force constant between the nearest-neighbor crystal planes, whose distance $a$ is of the order of $10^{-10}\,$m. In the limit of $q\ll 1/a$, which is valid for $\rC\gg10^{-9}\,$m \footnote{The relation in Eq.~\eqref{Lphonon}, shows that the wave number density is peaked near to $\sim\rC$.}, one can approximate the sine with its argument. Thus, we obtain
\bq \label{eq:gdfgf}
\omega_\text{\tiny L}(\q)\simeq v_\text{\tiny S}|\q|,
\eq
where $v_\text{\tiny S}$ is the speed of sound in the crystal (Eq.~\eqref{eq:gdfgf} is valid also for more general crystals). Using this expression, Eq.~\eqref{Lphonon} becomes
\bq\label{leffphonon}
\lambda_\text{\tiny eff}=\frac{4\lambda\rC^2\Omega_\text{\tiny c}^2}{3v_\text{\tiny S}^2}\!\left[\!\tfrac12-\frac{\rC^2\Omega_\text{\tiny c}^2}{v_\text{\tiny S}^2}+\sqrt{\pi}\frac{\rC^3\Omega_\text{\tiny c}^3}{v_\text{\tiny S}^3}e^{\tfrac{\rC^2\Omega_\text{\tiny c}^2}{v_\text{\tiny S}^2}}\!\!\erfc\left(\tfrac{\rC\Omega_\text{\tiny c}}{v_\text{\tiny S}}\right)
\!\!\right],
\eq
where $\erfc(x)=1-\erf(x)$.

Equations \eqref{dnsc}, \eqref{gammaXray} and \eqref{leffphonon} give the theoretical predictions for the cCSL model, which can be tested against experimental data. The resulting bounds   are reported in Fig.~\ref{fig.bounds} for different choices of $\Omega_\text{\tiny c}$.

\subsection{Details of the experimental setups}

Before discussing the results, be briefly describe the experimental setups, which we used in computing the bounds. 
\begin{figure*}[ht!]
\centering
\includegraphics[width=0.5\linewidth]{{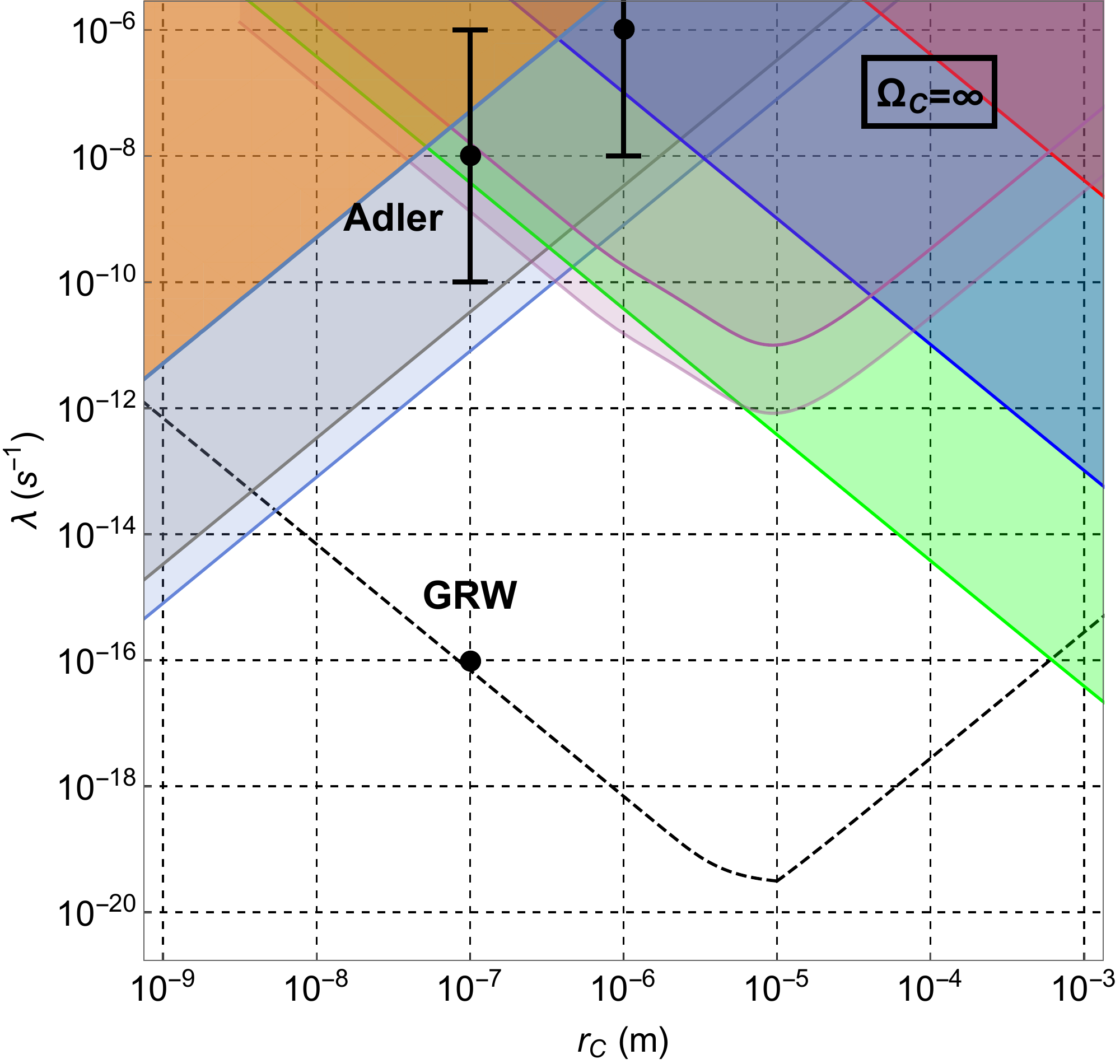}}\includegraphics[width=0.5\linewidth]{{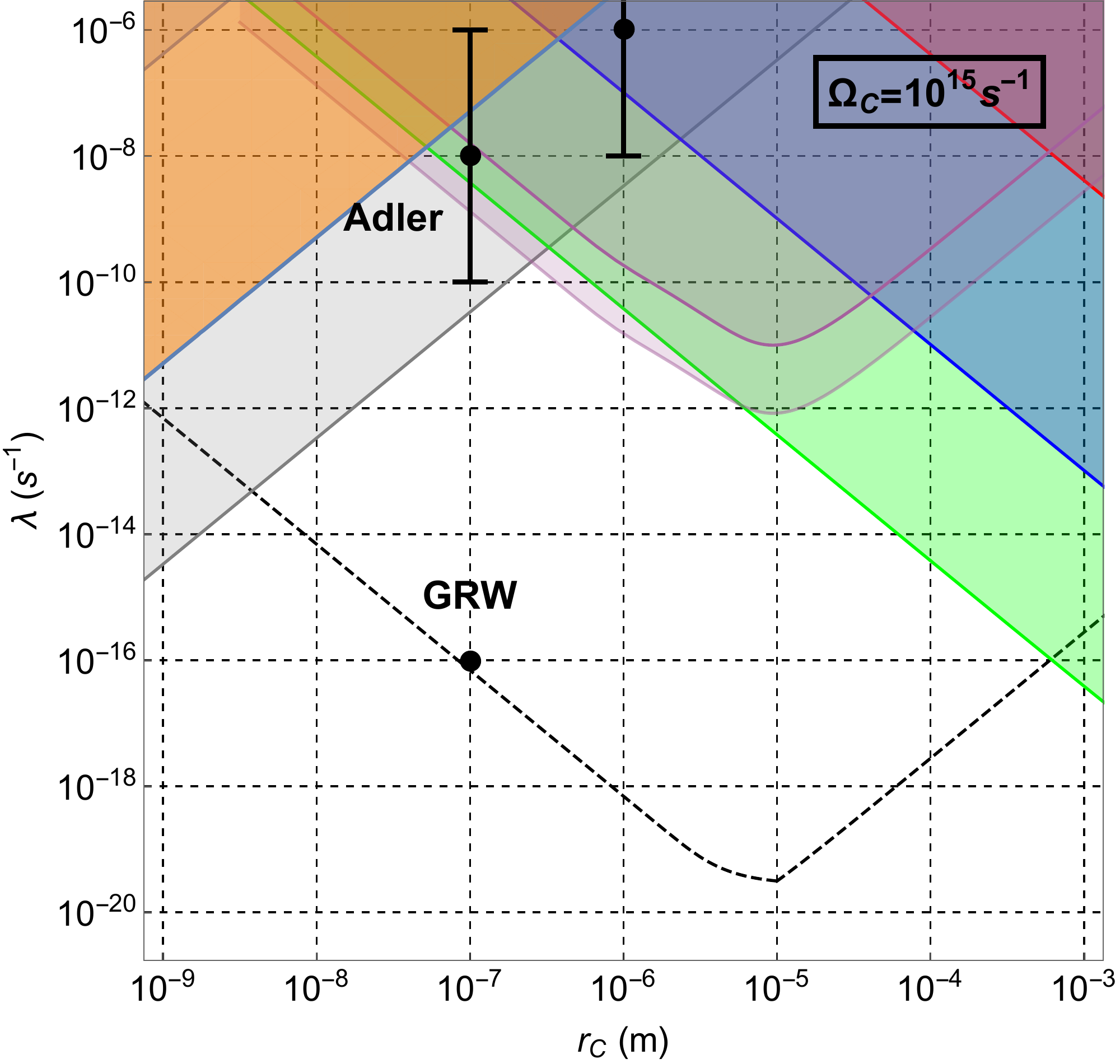}}
\includegraphics[width=0.5\linewidth]{{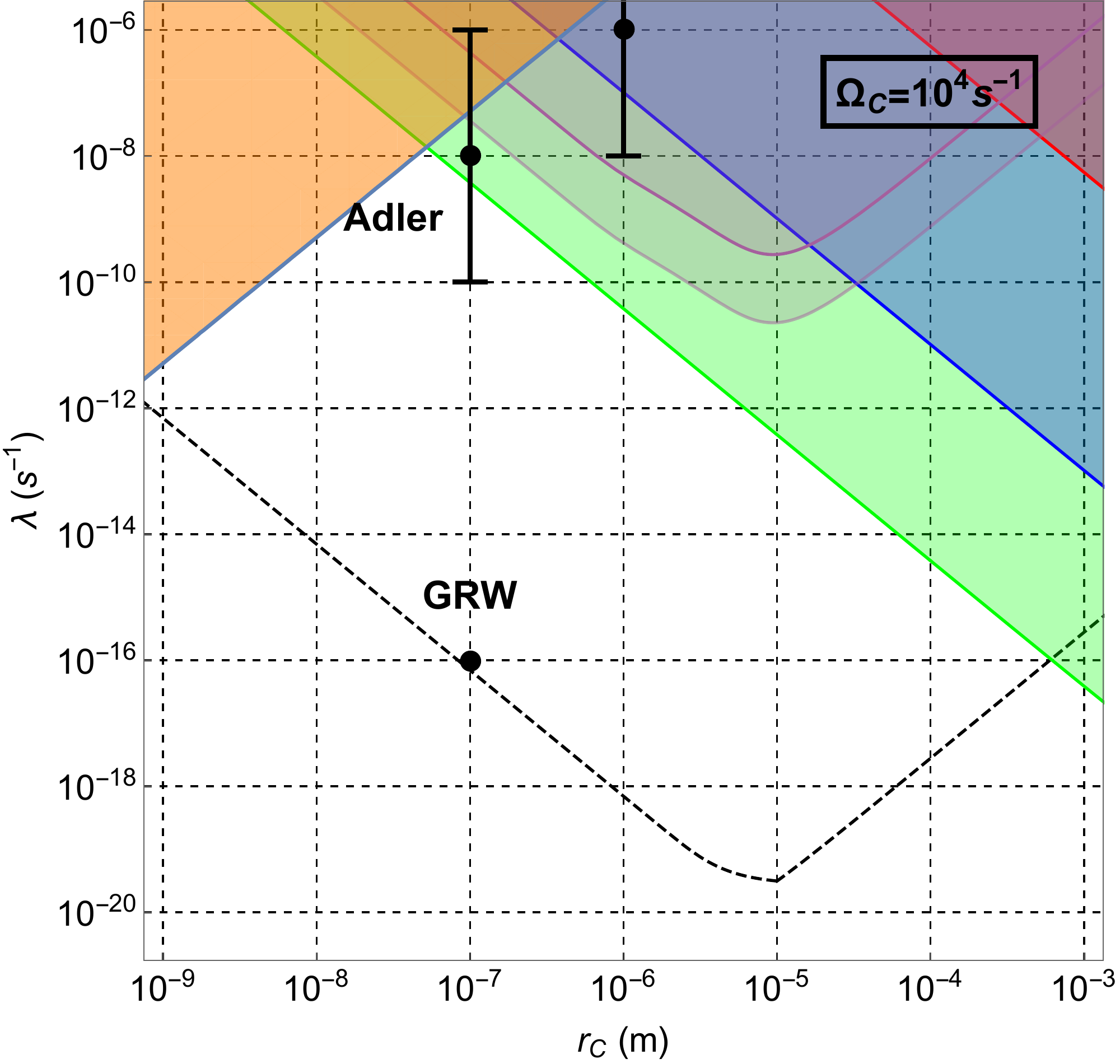}}\includegraphics[width=0.5\linewidth]{{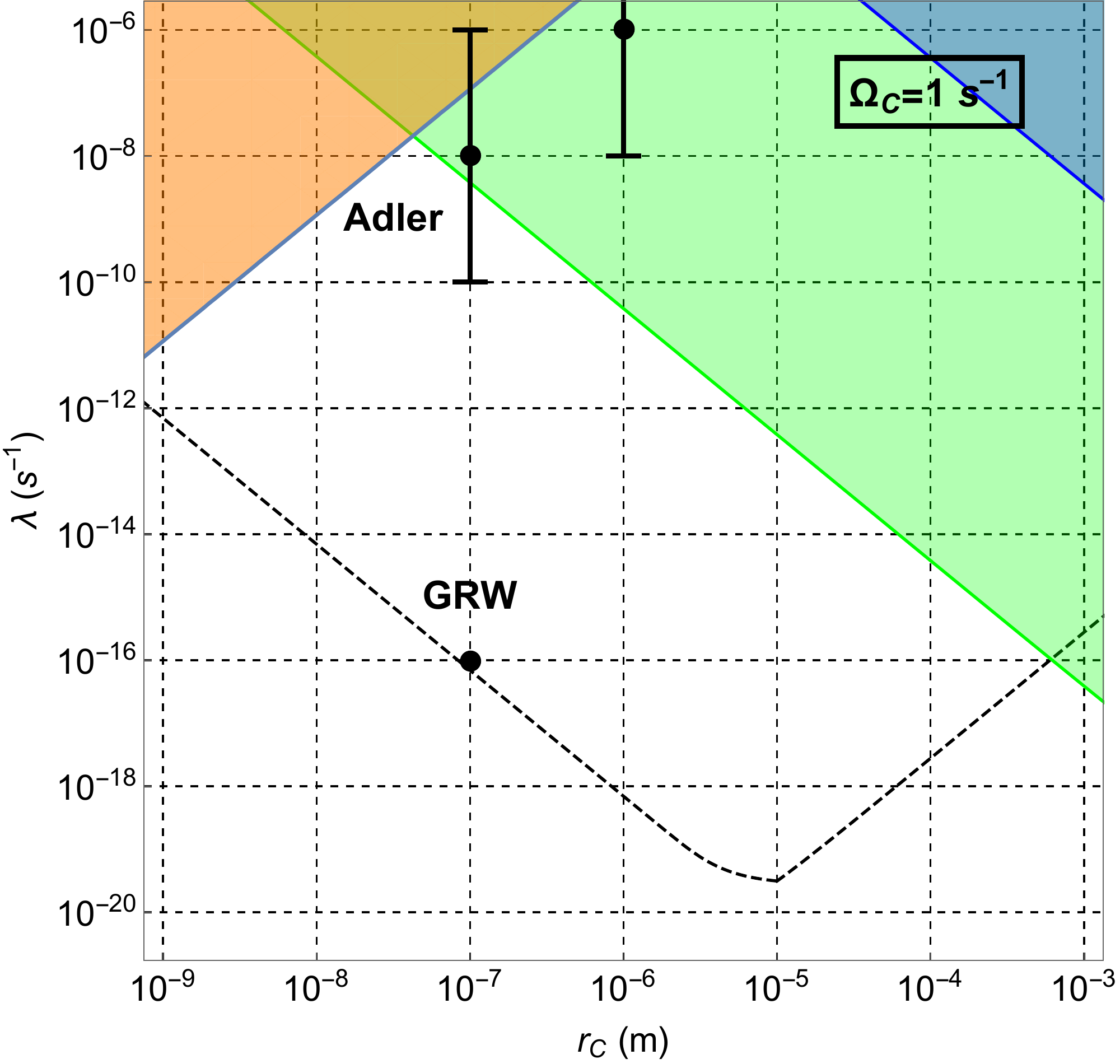}}
\caption{{(Color online) Upper and lower bounds on the cCSL parameters $\lambda$ and $\rC$ for three values of the frequency cutoff: $\Omega_\text{\tiny c}=+\infty$ (top left panel) corresponding to standard CSL, $\Omega_\text{\tiny c}=10^{15}\,$ Hz (top right), $\Omega_\text{\tiny c}=10^{4}\,$ Hz (bottom left), and $\Omega_\text{\tiny c}=10^{1}\,$ Hz (bottom right). Red, blue and green lines (and respective shaded regions): Upper bounds (and exclusion regions) from AURIGA, LIGO and LISA Pathfinder, respectively. Purple region: Upper bound from cantilever experiment. Light blue region: Upper bound from X-ray measurements. Orange and grey regions: Upper bound from cold atom experiment \cite{Kovachy:2015ab,Bilardello:2016aa} and  from bulk heating experiments \cite{Adler:2018aa}. The black dashed line shows the lower bound based on theoretical arguments \cite{Toros:2017aa}.}\label{fig.bounds}}
\end{figure*}

AURIGA is an aluminum cylindrical bar of length $3\,$m, radius $30\,$cm and mass $2300$ cooled down at 4.2\,K, whose resonant elongation is magnetically monitored at a frequency of $\omegam/2\pi\sim900\,$Hz \cite{Vinante:2006aa}. The contribution to the noise that can be atribuited to cCSL is $\DNSc\simeq1.4 \times10^{-22}\,$N$^2$/Hz at $\omega/2\pi=931\,$Hz, and the corresponding bound is reported in Fig.~\ref{fig.bounds} in red.

LIGO is a Michelson interferometer, whose two 4\,km arms are configured as a Fabry-Perot cavity \cite{Abbott:2016ab}. At each extreme of the two arms, a cylindrical silica mirror (density 2200\,kg/m$^3$, radius 17\,cm and length 20\,cm) are suspended and oscillate at a frequency below $1\,$Hz, while its noise is monitored in the $10-10^3\,$Hz band. The cCSL compatible contribution to the DNS is $\DNSc\simeq9\times 10^{-27}$\,N$^2$/Hz at $\omega/2\pi=30-35\,$Hz, which constrains the CSL parameters as reported in blue in Fig.~\ref{fig.bounds}. 

LISA Pathfinder is a space-based experiment which monitors the relative distance between two identical cubic masses (length 4.6\,cm, average distance 37.6\,cm, mass 1.928\,kg) at low frequencies \cite{Armano:2016aa,Armano:2018aa}. We can attribuite to cCSL a noise contribution of $\DNSc=3.15\times 10^{-30}\,$N$^2$/Hz just above the mHz regime. The corresponding bound is highlighted in green in Fig.~\ref{fig.bounds}.

Two cantilever experiments were performed with masses of $3.5\times10^{-13}\,$kg \cite{Usenko:2011aa,Vinante:2016aa} and $1.2\times10^{-10}\,$kg \cite{Vinante:2017aa}. Here, we focus on the second  experiment, which consists in a silica cantilever of dimension $450 \times 57\times2.5\,\mu$m$^3$ and stiffness 0.4\,N/m, to which is attached a ferromagnetical sphere of radius 15.5\,$\mu$m and density $7.43\times10^3\,$kg/m$^3$. The harmonic motion of the latter, which is characterized by a frequency $\omegam/2\pi=8174.01\,$Hz, is monitored with a SQUID. A CSL-like non-thermal contribution to the DNS was measured, taking the value $\DNSc=1.87\pm0.16\,$aN$^2$/Hz. Assuming that the measured extra noise is due to cCSL, the values of the cCSL parameters lie on the upper purple line in Fig.~\ref{fig.bounds}. Conversely, if such a noise can be attributed to standard sources, the experiment sets an upper bound corresponding with the lower purple line in figure.

The case of X-ray measurements~\cite{Fu:1997aa,Adler:2007ad,Adler:2013aa,Bassi:2014aa,Donadi:2014aa,Piscicchia:2017aa} is slightly different from previous ones because it relies on a different physical mechanism, the spontaneous emission of radiation rather than the Brownian motion. 
The expression for the photon emission rate in Eq.~\eqref{gammaXray} is compared with the experimental measure, which {gives $4\pi^2\epsilon_0c^3m_0^2 \omega\tfrac{\D\Gamma(\omega)}{\D\omega}/e^2\hbar\lesssim803\,$s$^{-1}$m$^{-2}$ \cite{Piscicchia:2017aa}. The corresponding upper bound is }reported in light blue in Fig.~\ref{fig.bounds}.

In  low temperature experiments~\cite{Pobell:2007aa} there is a residual heating of about $10^{-11}$ W/kg. This value should be compared with the energy rate in Eq.~\eqref{Erate}, where $\lambda_\text{\tiny eff}$ is estimated via Eq.~\eqref{leffphonon} with $v_\text{\tiny S}=3000\,$m/s (the  speed of sound in copper at low temperatures) \cite{Adler:2018aa}. The upper bound corresponding to these experiments is reported in grey in Fig.~\ref{fig.bounds}.

\section{Discussion and conclusion}
\label{sec.5}

The bounds reported in Fig.~\ref{fig.bounds} refer to four values of the cutoff frequency $\Omega_\text{\tiny c}$: 1, $10^4$, $10^{15}$\,$s^{-1}$ and $\infty$, the latter case corresponding to the standard CSL model.  

For $\Omega_\text{\tiny c}=10^{15}\,$s$^{-1}$, we notice the first change in the parameter space due to the colored extension of the model. The bound from  X-ray measurements becomes weaker: the reason is that frequency of the X-rays ($\omega \sim 10^{19}\,$\,s$^{-1}$), at which the collapse noise is sampled,  exceeds the cutoff frequency. The next to vanish is the bound on phonon excitations, which samples the noise at frequencies $\sim 10^{11}\,$s$^{-1}$.
Similarly, the bound from the cantilever experiment weakens for $\Omega_\text{\tiny C}\leq10^4\,$s$^{-1}$, as shown in the last panel of Fig.~\ref{fig.bounds}. The same happens for the bounds from AURIGA, LIGO and LISA Pathfinder when the cutoff frequency takes values $\Omega_\text{\tiny c}\lesssim10^3\,$s$^{-1}$, $10^2\,$s$^{-1}$  and $10^{-2}\,$s$^{-1}$ respectively. 
Eventually, for a  cutoff at $\Omega_\text{\tiny c}=10^{1}\,$s$^{-1}$, the only relevant bounds are those coming from cold atom experiments and LISA Pathfinder.

If one assumes that cCSL has a cosmological origin, which is reasonable considering the universality of the noise, a hint on the value of $\Omega_\text{\tiny c}$ could come from the behaviour of cosmological fields \cite{Bassi:2010aa}. If one takes the Cosmic Microwave Background (CMB) radiation, or the relic neutrino background, one has  $\Omega_\text{\tiny c}\sim10^{12}\,$s$^{-1}$ \cite{Grishchuk:2010aa}. For such a value, the cCSL bound from X-rays is completely washed away. Thus, the bound from bulk heating effects \cite{Adler:2018aa,Bahrami:2018aa} becomes the strongest. For $\Omega_\text{\tiny c}=10^{11}\,$s$^{-1}$ also the latter vanishes and the cold atom bound \cite{Kovachy:2015ab,Bilardello:2016aa} prevails for $\rC<10^{-7}\,$m \footnote{We believe that the CSL contribution to the position diffusion of the cold atom cloud in the colored case, described by Eq.~(55) of \cite{Bilardello:2016aa}, should be substituted with
$
\frac{3\lambda A^2 \hbar^2}{2m^2\rC^2}\left[	\frac{t^3}{2}-\frac{t^2\tau}{2}+\tau^2\left(	\tau-(t+\tau)e^{-t/\tau}\right)\right]
$, where $A$ is the atomic number, $m$ is the mass of the sigle atom and $\tau=\Omega_\text{\tiny c}^{-1}$. We used such an expression to draw the corresponding bounds in Fig.~\ref{fig.bounds}.}, while for $\rC>10^{-7}\,$m the strongest  bounds are provided by cantilever experiments and LISA Pathfinder.

We can conclude that  low frequency, purely mechanical experiments provide the most robust bounds on the CSL parameters. These are resistant against changes in the spectrum of the noise, unless of course for some reason the low frequency part of the noise spectrum is suppressed for some reason, which cannot be identified at the present stage. However this possibility would compromise the reduction process, which requires non-vanishing low frequency components \cite{Adler:2007aa,Adler:2008aa}.

\acknowledgments
\noindent  The authors wish to thank S L Adler for useful comments on a preliminary draft of the paper.
MC and AB acknowledge financial support from the  H2020 FET project TEQ (grant n.~766900). AB acknowledges financial support from the University of Trieste (FRA 2016), INFN and the COST Action  QTSpace (CA15220), and hospitality from the IAS Princeton, where part of this work was carried out and partial financial support from FQXi.
 LF acknowledges funding from the Royal Society under the Newton International Fellowship No. NF170345.

\subsection*{Author contribution statement}

\noindent M.C.~and A.B.~conceived the presented idea. M.C.~and L.F.~developed the theory and derived the experimental bounds. A.B.~coordinated the research project. All authors contributed to the writing of the manuscript.

\end{document}